\documentclass[pra,twocolumn,showpacs,nobibnotes,aps]{revtex4-1}

\usepackage[utf8]{inputenc}
\usepackage{amsmath}
\usepackage{graphicx}
\usepackage{hyperref}

\newcommand{\ket}[1]{| #1 \rangle}

\newcommand{\braket}[2]{\langle #1 | #2 \rangle}
\newcommand{\opbraket}[3]{\left\langle #1 \right|#2 \left| #3 \right\rangle}
\newcommand{\bigket}[1]{\left| #1 \right\rangle}

\newcommand{\Uint}{U_{\rm int}}
\newcommand{\Uintopt}{U_{\rm int}^{\rm opt}}
\newcommand{\Vlift}{V_{\rm lift}}
\newcommand{\psii}{\psi_{\rm i}}
\newcommand{\psit}{\psi_{\rm t}}
\newcommand{\mlvl}{m_{\rm L}}
\newcommand{\mRb}{m_{\rm Rb}}

\newcommand{\fref}[1]{Fig.~\ref{#1}}
\newcommand{\sref}[1]{Sec.~\ref{#1}}
\newcommand{\eref}[1]{Eq.~(\ref{#1})}

\newcommand{\Eref}[1]{Equation~(\ref{#1})}

\usepackage{color}

\begin{document}

\preprint{APS/123-QED}

\title{A robust boson dispenser: \\ Quantum state preparation in interacting many-particle systems}

\author{Irina Reshodko}
\email{irina.reshodko@oist.jp}
\affiliation{OIST Graduate University, 904-0495 Okinawa, Japan}

\author{Albert Benseny}
\affiliation{OIST Graduate University, 904-0495 Okinawa, Japan}

\author{Thomas Busch}
\affiliation{OIST Graduate University, 904-0495 Okinawa, Japan}

\date{\today}

\begin{abstract}
We present a technique to control the spatial state of a small cloud of interacting particles at low temperatures with almost perfect fidelity using spatial adiabatic passage.
To achieve this, the resonant trap energies of the system are engineered in such a way that a single, well-defined eigenstate connects the initial and desired states and is isolated from the rest of the spectrum. We apply this procedure to the task of separating a pre-defined number of particles from an initial cloud and
 show that it can be implemented in radio-frequency traps using experimentally realistic parameters.
\end{abstract}


\maketitle

\section{\label{sec:intro}Introduction}

Small samples of ultra-cold atoms trapped in external potentials are shaping up to become paradigmatic systems for exploring the fundamental building blocks of quantum many-body dynamics \cite{Stroescu2016,Wenz2013, Greiner2001,Garcia-March2013, Garcia-March2013a}.
While in the weakly interacting regime samples with more than five atoms are well described by a mean-field approach~\cite{Stoof2009, Wenz2013}, strongly interacting systems have been shown to allow for the creation of highly correlated quantum many-body states~\cite{Haldane1988, Islam2015a}.
Understanding and controlling interactions and many-particle dynamics is therefore crucial for accessing a larger part of these systems' Hilbert space.

One important ingredient in this quest is the development of high-fidelity quantum engineering techniques. This, however, is a non-trivial task, due to the large number of degrees of freedom present in many-particle systems, which make it hard to follow or reach specific states. It is therefore sensible to start the development with systems with only a small number of particles and later generalize the developed tools to larger systems.

One of techniques which allow for high-fidelity state preparation in external potentials is 
spatial adiabatic passage (SAP)~\cite{SAPreview2016}. It is an analogue to the well-known STIRAP technique in atomic physics~\cite{Bergmann1998,Bergmann2015,Vitanov2016} and utilizes the existence of a specific ``dark'' eigenstate to coherently transfer single particles between two localized spatial states~\cite{Eckert2004}.
Compared to STIRAP, the SAP setting can possess a larger variety of degrees of freedom, which has in recent years allowed to extend the technique to multiple dimensions~\cite{ORiordan2013,McEndoo2014,Polo2015} , angular momentum states~\cite{McEndoo2010,Polo2016}, non-linear systems~\cite{Graefe2006,Rab2008,ORiordan2013,Rubio2016} and interacting particles~\cite{Benseny2010,Bradly2012,Benseny2016,Gajdacz2011, Gajdacz2014}.

In this work we will extend the previous developments on interacting systems and show that the typical control that exists in ultracold atom experiments can be used to devise techniques based on SAP for the engineering of specific many particle states. 
For this we will investigate the possibility of creating a single particle source from two- and three-particle samples.
We will also discuss the generalization of the proposed method to engineer the separation of a single particle from an arbitrary initial number of interacting particles, as well as the separation of an arbitrary, but well defined, number of particles from the initial system.  To show that our proposal is realistic, we also discuss an  implementation of the suggested protocol using radio-frequency traps.

Our mansucript is structured as follows.
In order to understand the proposed protocols, we first briefly review the main ideas of the single-particle and two-particle SAP protocols in \sref{sec:1p2pSAP}.
Then, in \sref{sec:part_sep}, we present the modified SAP protocols designed for the separation of a specified number of particles from an atomic cloud and discuss their limits.
To show that our ideas are realistic, we present in \sref{sec:rf_background}  a study of a potential implementation of the particle separation protocol using radio-frequency traps. Finally, we conclude in \sref{sec:conclusions}.

\section{Single-particle and two-particle SAP}
\label{sec:1p2pSAP}

\begin{figure}[tb]
	\includegraphics[width=0.8\linewidth]{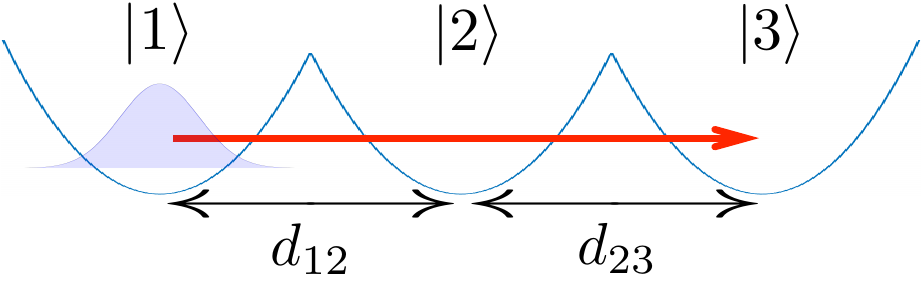}
	\caption{Schematic of the SAP setup using a triple harmonic trap system.  The ground states of the left, middle and the right trap are given by $\ket{1}=\ket{1~0~0}$, $\ket{2}=\ket{0~1~0}$, and $\ket{3}=\ket{0~0~1}$, respectively.
	The distances  $d_{12}$  and $d_{23}$ between the traps can be changed independently and 
	the goal is to achieve a high fidelity transfer of a particle from the left trap to the right one. 
	}
\label{fig:SAP}
\end{figure}

The fundamental principle behind SAP can be illustrated by considering a one-dimensional model in which a single particle is trapped in an external potential consisting of three truncated harmonic traps (see \fref{fig:SAP})~\cite{Eckert2004}
\begin{equation}
	V(x) = \frac{1}{2}m\omega^2\min\left[(x+d_{12})^2, x^2, (x-d_{23})^2\right].
    \label{eq:truncqho}
\end{equation}
Here $d_{12}$ and $d_{23}$ are the distances between the minima of the left and middle traps  and the middle and right traps, respectively, $m$ is the mass of the particle and $\omega$ is the trapping frequency, which is taken to be identical for all three traps.
Assuming that the particle is initially in the center-of-mass ground state of the left trap and that the evolution is carried out adiabatically, this system can be modelled by considering only the ground states of the three traps, $\ket{1}, \ket{2}$ and $\ket{3}$. 
The Hamiltonian of such three-level system can then be written as
\begin{equation}
\label{eq:H3}
	\hat{H}(t) = \hbar
					\begin{pmatrix}
						0&\Omega_{12}(t)&0\\
						\Omega_{12}(t)&0&\Omega_{23}(t)\\
						0&\Omega_{23}(t)&0
					\end{pmatrix},
\end{equation}
where the $\Omega_{ij}$ are the coupling frequencies between the states $\ket{i}$ and $\ket{j}$ which depend on the distance between the traps $d_{ij}$ for $i,j = 1, 2, 3$. 
Diagonalizing this Hamiltonian gives one eigenstate, the so-called {\it dark state}, with zero eigenvalue, which only has contributions from the traps on the left and on the right

\begin{equation}
\label{eq:dstate}
	\ket{D(\theta)} = \cos \theta \ket{1} - \sin \theta \ket{3},\quad \tan \theta = \frac{\Omega_{12}}{\Omega_{23}}.
\end{equation}
SAP then consists in following the dark state and transferring a particle from the left trap to the right by adiabatically changing $\theta$ from $0$ to $\pi/2$.
This requires to change the relative coupling strengths between the traps, which can be done by approaching and separating individual pairs. 
The movement sequence is famously counter-intuitive because the right and the middle traps, which are both empty, approach each other before the left trap starts moving. Since the process does not depend on the exact form of the positioning sequence, it is robust to experimental uncertainties.

The introduction of interactions to the system yields a loss of resonance in the tunnelling process which, in principle, requires careful and time-dependent trapping potential adjustments~\cite{ORiordan2013,Graefe2006,Rubio2016}.
However, it was recently shown that these adjustments are not necessary 
in few-particle systems for a large range of interaction strengths~\cite{Benseny2016}.
The Hamiltonian for $N$ ultra-cold bosons of mass $m$ in one dimension can be written as

\begin{equation}
    \label{eq:ham}
    \hat{H} = \sum_{j=1}^N\left[-\frac{\hbar^2}{2m}\frac{\partial^2}{\partial x_j^2} + V(x_j)\right] + g\sum_{k > j}\delta(x_j-x_k) ,
\end{equation}
where $V(x)$ is again given by~\eref{eq:truncqho} and the interaction between the particles is assumed to be point-like with strength $g$~\cite{Lieb1963,Olshanii98}. 
To understand the underlying principles of how SAP works in interacting systems, which also are the key to understanding the techniques in the following section, we will briefly review the two-particle case here. It is described by the explicit Hamiltonian
\begin{equation}		
    \label{eq:ham2p}
    \hat{H} = -\frac{\hbar^2}{2m}\frac{\partial^2}{\partial x_1^2} -\frac{\hbar^2}{2m}\frac{\partial^2}{\partial x_2^2} + V(x_1) +V(x_2) + g\delta(x_1-x_2),
\end{equation}
and the strength $g$ directly relates to the spectrum for two particles in a harmonic trap as~\cite{Busch98}
\begin{equation}
	g =- \frac{2\sqrt{2} \Gamma(1-E_g/2)}{\Gamma((1-E_g)/2)},
\end{equation}
where $\Gamma(E)$ is the gamma function. 
Thus, we can define an interaction energy $\Uint$ as
\begin{equation}
\Uint = E_g - 2 E_0,
\end{equation}
where $E_0 = \hbar \omega/2$ is the (single-particle) harmonic oscillator ground state energy and $E_g$ is the two-particle ground state energy.
In this and the following section we will use natural units where $\hbar=m=\omega=1$.

The existence of a  range of intermediate interaction strengths where high fidelity SAP transfer can take place can then be understood from the band structure of the Hamiltonian spectrum~\cite{Benseny2016}.
The lowest energy band has energies around 1 and contains states where the two atoms are in the ground states of different traps.
The second band, which lies around energy values between 1 and 2 (depending on the interaction strength), corresponds to states where both particles are in the same trap, and contains a dark state similar to the one in \eref{eq:dstate} which allows for the transport of the particle pair.
Higher bands correspond to states where at least one of the atoms is in an excited trap state.
For those intermediate interaction strengths where the second band remains isolated from the other two bands, the dark state can be adiabatically followed and high-fidelity SAP transport can be achieved~\cite{Benseny2016}.

In this regime of intermediate interactions, the two particles are repulsively bound and effectively behave like a single particle~\cite{Winkler2006}.
It is therefore clear that all single particle protocols transfer directly to this situation, and in particular it is straightforward to engineer a two particle NOON state by changing $\theta$ from $0$ to $\frac{\pi}{4}$. 
This leaves the system in the state $\frac{1}{\sqrt{2}}(\ket{2~0~0}  - \ket{0~0~2} )$, with $\ket{2~0~0}$ and $\ket{0~0~2}$ denoting a state with two particles in the left and in the right trap respectively.
NOON states are maximally entangled and are important in quantum engineering and quantum metrology~\cite{Lee2002,Schloss2016}, as they allow for phase measurements that can reach the fundamental Heisenberg limit~\cite{Boto2000}.

\section{Particle separation}
\label{sec:part_sep}

In the following we will discuss a process based on SAP which is not a straightforward generalization of a single particle protocol, but which allows to  split an initial many-particle state in a controlled manner. 

\subsection{Two-particle case}
\label{sec:2part_full}

To demonstrate the principle of the process, let us initially assume that the left trap contains two particles only, and that the target state of the process has one particle in the left and the other in the right trap
\begin{equation}
	\ket{\psi_\textrm{i}}=\ket{2~0~0} \ \rightarrow \ \ket{\psi_\text{f}}= \ket{1~0~1}.
	\label{eq:process}
\end{equation}

\begin{figure}
\begin{center}
\includegraphics[width=0.95\linewidth]{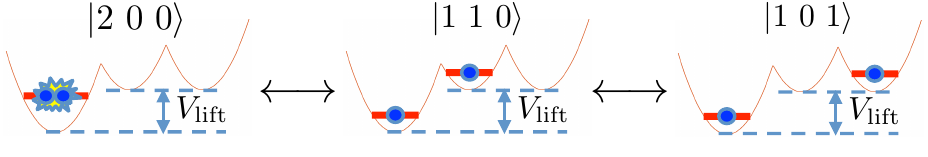}
\caption{
Schematic of the three-level model for particle separation.
Simultaneous lift of the right and the middle harmonic traps by $\Vlift=\Uint$ makes the three states depicted resonant.
}
\label{fig:sq_lift}
\end{center}
\end{figure}

Due to the atomic interactions, the initial and the final state have different energies and are in different energy bands.
To make the desired coupling possible, it will therefore be necessary to match the energies and compensate for the absence of the interaction energy in the final state by adjusting the energies of the traps. 
This can be done in a time-independent manner by raising the energies of the middle and the right traps by $V_\text{lift} = \Uint$
(or lowering in the case of attractive interactions),
which ensures resonance between states $\ket{2~0~0}$, $\ket{1~0~1}$, and $\ket{1~1~0}$. This also energetically separates them from all other states, making the system effectively a three-level system, analogous to \eref{eq:H3} (see \fref{fig:sq_lift}). 
A dark-like energy eigenstate, which involves only the initial and target states can then be found and the counter-intuitive positioning sequence leads to the desired particle separation. 

\begin{figure}
\begin{center}
\includegraphics[width=0.9\linewidth]{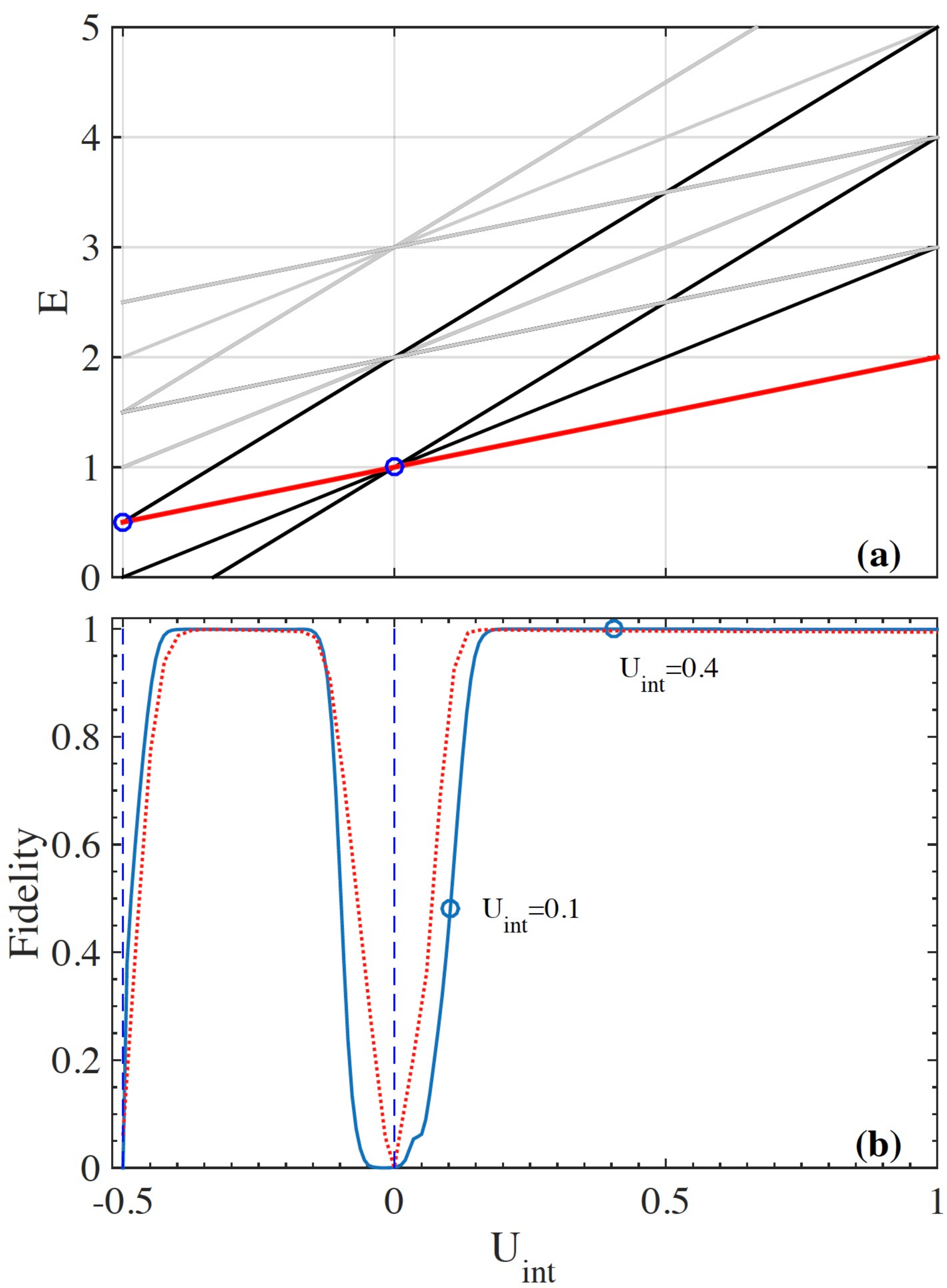}
\caption{\textbf{(a)} Bose--Hubbard spectrum of the two-particle Fock states in the three trap system with only the lowest two energy levels in each trap considered and $\Vlift = \Uint$.
The three degenerate states $\ket{2~0~0}, \ket{1~1~0}$ and $\ket{1~0~1}$ are in the band colored in red and
additional degeneracies can be seen to appear at $\Uint = 0$ and $\Uint = -{1}/{2}$.
\textbf{(b)} Fidelities of the particle separation process as a function of the interaction energy, obtained using the full Hamiltonian time evolution (solid blue line) and BH model (dashed red line).
Degeneracies of the spectrum appear at points marked as vertical dashed blue lines.
}
\label{fig:es_lift_2}
\end{center}
\end{figure}
\begin{figure}
\begin{center}
\includegraphics[width=0.95\linewidth]{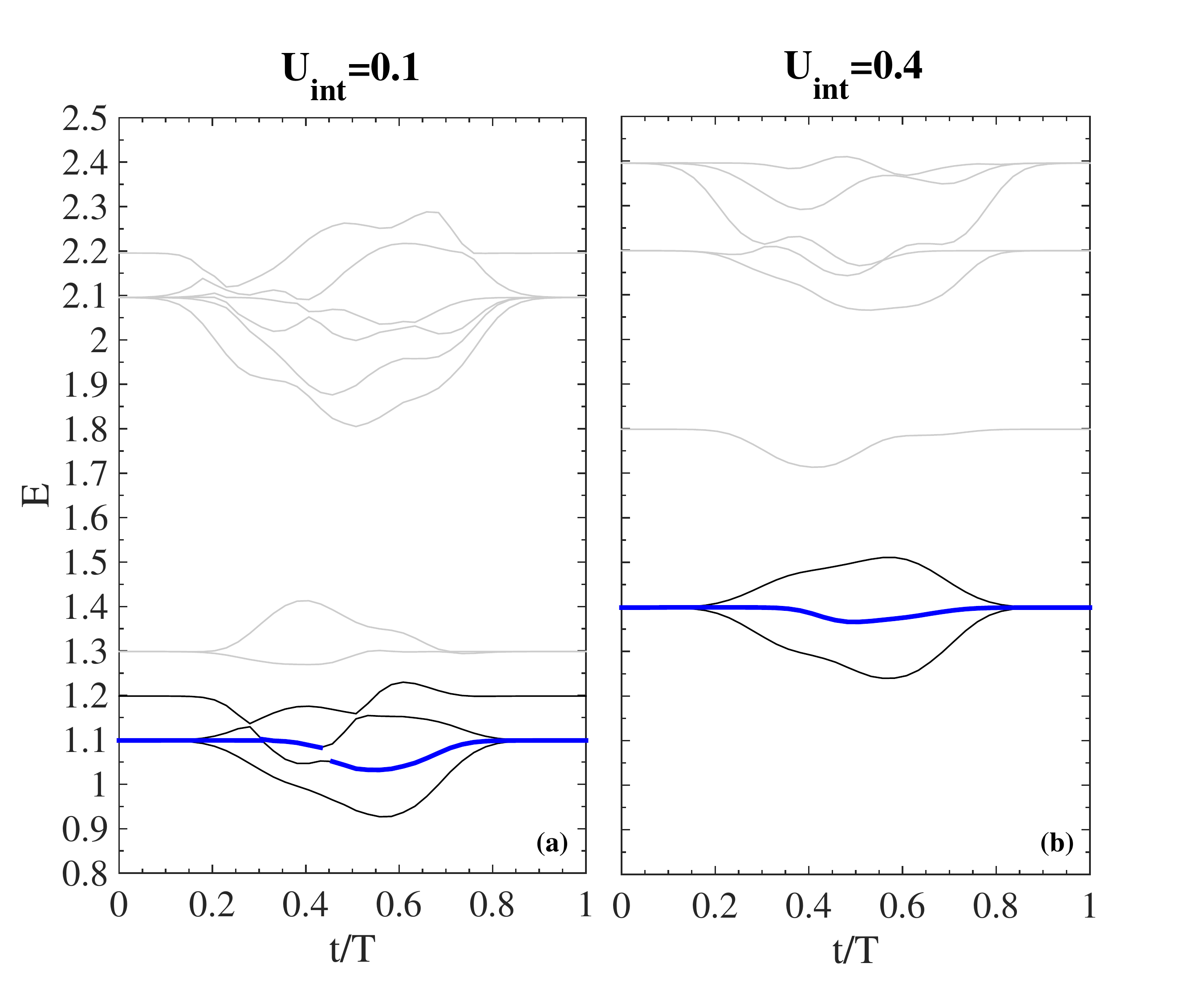}
\caption{Evolution of the energy spectrum  during the  particle separation process for
(a) $\Uint = 0.1$ and 
(b) $\Uint = 0.4$. 
The dark-like state is highlighted in blue.
}
\label{fig:es_lift_1}
\end{center}
\end{figure}
 
To confirm that the splitting process works as expected we will in the following  simulate the above system first in an ideal setting. 
Using the full Hamiltonian given in \eref{eq:ham2p}, we calculate the fidelity $F = \left| \braket{\psi_\text{f}}{\psi_\text{T}} \right|^2$ of the process, which is shown in \fref{fig:es_lift_2}(b) for $\Uint\in\left[-\frac{1}{2}, 1\right]$  (solid blue line).
Although there is a prominent dip in the weakly-interacting regime, the process can be seen to result in high fidelities over a wide range of interaction strengths.
This is easy to understand in the limit of infinitely repulsive interactions (the Tonks--Girardeau limit, $\Uint=1$), where the bosonic particles can be described as non-interacting fermions \cite{Girardeau2001}.
The system can then be thought of as forming a {\it Fermi sea} in the harmonic trap, and by choosing $V_\text{lift}=1$  only the particle at the Fermi edge can tunnel.
It is therefore natural to expect high-fidelity particle separation in this case. 

The drop of fidelity in the weakly-interacting regime can be understood by looking at   
the time-dependent spectrum of the Hamiltonian~\eqref{eq:ham2p}, which we show  in \fref{fig:es_lift_1} for two different values of $\Uint$ (corresponding to the points indicated in \fref{fig:es_lift_2}(b)).
One can see that for weak interactions ($U_\text{int}=0.1$) the lowest band, which contains the dark state, overlaps with the next higher lying one and therefore level crossings lead to the degraded fidelity. 
For stronger interactions ($U_\text{int}=0.4$), the band-overlap vanishes and following the dark state becomes possible.
The drop of fidelity for $\Uint=-\frac{1}{2}$ will be discussed below, in \sref{sec:BH}. 

In the following sections we will extend the separation idea discussed above to systems with larger numbers of particles, $N$.
However, since the resources required to diagonalize and numerically integrate the Schr\"odinger equation using the full Hamiltonian~\eqref{eq:ham} scale exponentially with $N$, we will in the next section introduce a Bose--Hubbard (BH) model for this three-trap system.
To establish correspondence, we will first compare the above results for the  two-particle case to the two-particle BH model and then use the BH model to simulate the three-particle case.

\subsection{Bose--Hubbard model}
\label{sec:BH}

Let us assume a system of $N$ bosons distributed over three traps, which are lifted by energy values $V_1$, $V_2$ and $V_3$  (counted from left to right) with $\mlvl$ vibrational states in each trap.
Such a system contains ${N+3\mlvl-1}\choose{3\mlvl-1}$ Fock states 
and we associate each state with a matrix $\{n_{ji}\}$, where $n_{ji}$ is the number of particles in the $j$-th energy level of the $i$-th trap
\begin{equation}
\ket{\{n_{ji}\}}=\bigket{\begin{smallmatrix}
						n_{01}&n_{02}&n_{03}\\
						&\ldots&\\
						\ldots&n_{ji}&\ldots\\
						&\ldots&\\
						n_{(\mlvl-1)1}&n_{(\mlvl-1)2}&n_{(\mlvl-1)3}\\
\end{smallmatrix}}.
\end{equation}
Summing over all states and occupation numbers gives the overall number of particles, $\sum_{i=1}^{3}\sum_{j=0}^{\mlvl-1}n_{ji}=N$.
For states in which only the lowest  band is occupied we use the more intuitive notation 
\begin{equation}
\ket{\{n_{0i}\}}=\ket{n_{01}~n_{02}~n_{03}}.
\end{equation}
The BH Hamiltonian for this system can then be written as
\begin{align}
	\hat{H}_\textrm{BH} &=
	\hbar\omega\sum_{j=0}^{\mlvl-1}\left(j+\frac{1}{2}\right)\hat{N}^\textrm{level}_j
	+ \sum_{i=1}^{3}V_i \hat{N}^\textrm{trap}_i
	\nonumber\\ 
	&\qquad+ \frac{\Uint}{2}\sum_{i=1}^{3}\hat{N}^\textrm{trap}_i\left(\hat{N}^\textrm{trap}_i-1\right) + H_\text{tunnel}, 
\label{eq:BHHam}
\end{align}
where the $a_{ij}$ are the annihilation operators for a boson in the $j$-th level of trap $i$ and the
$
\hat{n}_{ji} = \hat{a}^\dagger_{ji}\hat{a}_{ji}
$
are their associated particle number operators. 
The total number of particles in the $i$-th trap is therefore
\begin{equation}
\hat{N}^\textrm{trap}_i = \sum_{j=0}^{\mlvl-1}\hat{n}_{ji},
\end{equation}
with corresponding eigenvalues $N^\textrm{trap}_i$ and the total number of particles in the $j$-th band is 
\begin{equation}
\hat{N}^\textrm{level}_j=\sum_{i=1}^{3}\hat{n}_{ji}.
\end{equation}
with eigenvalues $N^\textrm{level}_j$.
The first two terms of \eref{eq:BHHam} correspond to the single particle eigenenergies of the atoms in their respective levels, the third term describes the particle-particle interactions, and the last term accounts for all events where $p$ particles tunnel between two adjacent traps (the details on how to calculate the respective tunnel couplings are given in the Appendix).

When the traps are far apart and the tunneling couplings are negligible it is sufficient to consider only the first three terms of the Hamiltonian in \eref{eq:BHHam}, which has the Fock states $\ket{\{n_{ji}\}}$ as its eigenstates.
Their associated energies, $E_\textrm{total}(\{n_{ji}\})$, 
then depend only on the interaction $\Uint$ and potential lifts $V_i$.
In particular, for states in the lowest band, $\ket{n_{01}~n_{02}~n_{03}}$, $E_\textrm{total}$ reduces to
\begin{align}
	E_\textrm{total}(\{n_{0i}\}) &= \frac{N}{2}\hbar\omega+ \sum_{i=1}^{3}V_i n_{0i} + \frac{\Uint}{2}\sum_{i=1}^{3}n_{0i}(n_{0i}-1).
	\label{eq:TotEnergyFockLowest}
\end{align} 
The resonance condition between two Fock states, $\ket{\{n_{ji}\}}$ and $\ket{\{n^\prime_{ji}\}}$, can be written as
\begin{equation}
E_\textrm{total}(\{n_{ji}\}) = E_\textrm{total}(\{n^\prime_{ji}\}) ,
\label{eq:degen}
\end{equation}
which can be used to find the appropriate trap lifts to create the SAP triplet of resonant states ($\ket{2~0~0}$, $\ket{1~1~0}$, and $\ket{1~0~1}$) for the particle separation protocol used above
\begin{equation}
V_1=0; \quad \Vlift=V_2=V_3=\Uint.
\end{equation}

The fidelities for the SAP separation process obtained from this BH model can be seen in \fref{fig:es_lift_2}(b) to be very similar to the ones obtained using the full Hamiltonian. Both approaches show large plateaus of high fidelity with drops around $\Uint = 0$ and $-\frac{1}{2}$.
This can be understood in the BH model by examining the spectrum of the two-particle Fock states, considering only the lowest two Bloch bands ($\mlvl = 2$), shown in \fref{fig:es_lift_2}(a). 
The highlighted red line corresponds to the energy of the degenerate SAP triplet and crossings between this band and other Fock states appear exactly at $\Uint=0$ and $\Uint=-1/2$.
The drops in fidelity around these values can therefore be attributed to these level crossings and,
in particular, for $\Uint = -1/2$, the band that is crossed corresponds to the one containing states where one particle is in the ground state of the left trap and the other is in the first excited state of the middle or the right trap.

This demonstrates that the BH model reproduces the main features of the full model and we will be using it in the following to design and simulate particle separation processes in systems with larger particle numbers.

\subsection{$N$-particle case}
\label{sec:npart}

\begin{figure}
\begin{center}
{\qquad\small   $\ket{\psit} =\ket{1~0~2}$\qquad\qquad \qquad $\ket{\psit} = \ket{2~0~1}$}
\includegraphics[width=\linewidth]{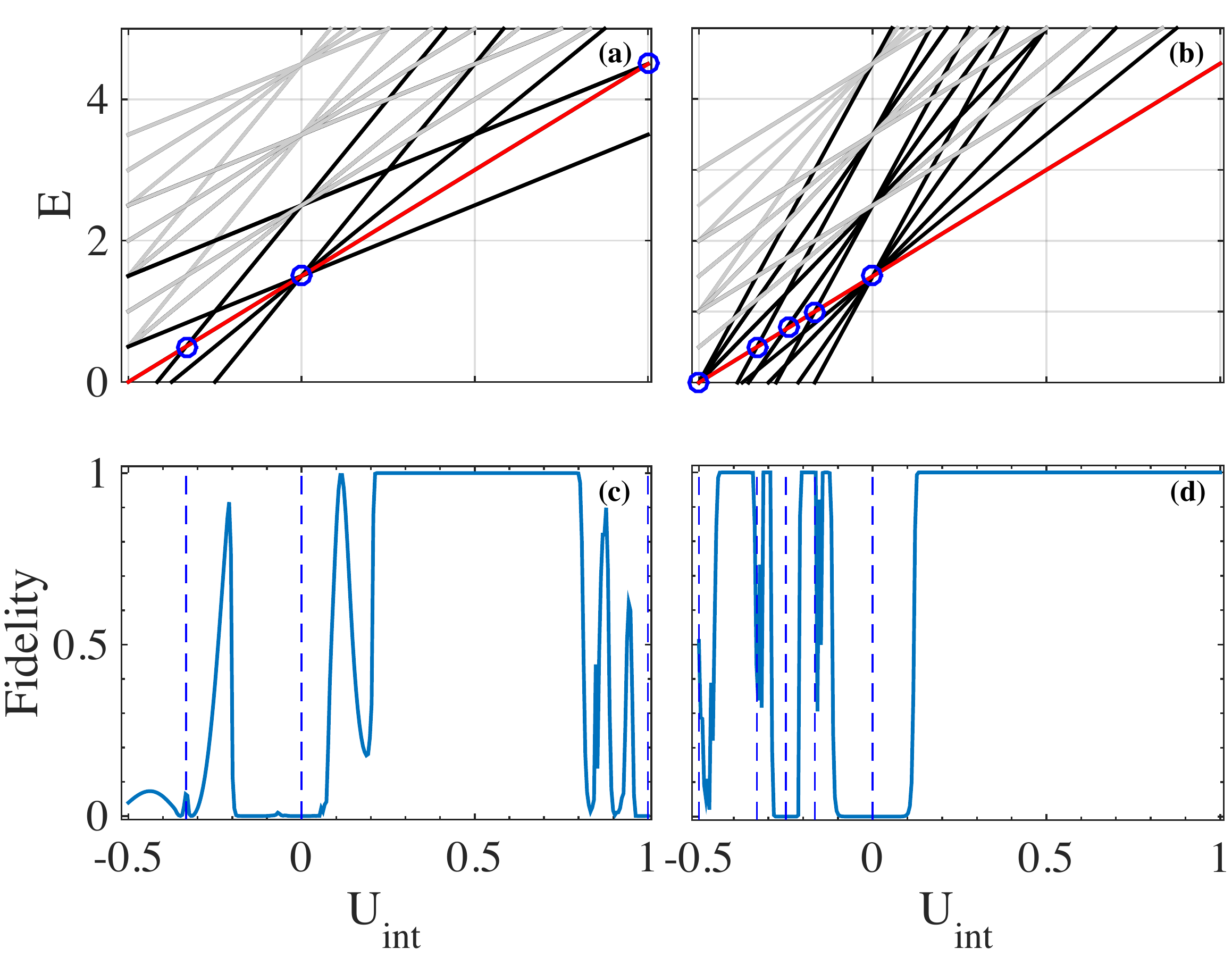}
\caption{
(a) Energy spectrum of three-particle Fock states in the BH model for $\mlvl = 2$ for the target state $\ket{1~0~2}$ (and $\Vlift = \Uint$). 
The energy of the SAP triplet is highlighted in red.
(c) Corresponding particle separation fidelities. 
(b,d) are the same as (a,c) but for $\ket{2~0~1}$ (and $\Vlift = 2\Uint$).
The circles in the top row and vertical dashed lines in the bottom row indicate the positions where level crossings between the SAP triplet and other bands exist.
}
\label{fig:es_3qho_3P_FockEn}
\end{center}
\end{figure}

Starting with a cloud of $N$ particles initially located in the left trap, we will show in this section that it is possible to separate exactly $M$ particles out of it.
While the preparation of such an initial state is by today still experimentally challenging, recent progress in this direction has shown that this can be done for a wide range of particle numbers~\cite{Stroescu2016}.
We will consider first the case where after the SAP dynamics exactly $M$ particles remain in the left trap, and later consider the case where exactly $M$ particles are separated into the right trap. 
The differences between these two cases will be explained below.

For the first case the initial and the target state are given by
\begin{equation}
\ket{\psii} = \ket{N~0~0}
\ \rightarrow \
\ket{\psit} = \ket{M~0~(N-M)} ,
\end{equation}
and the degeneracy conditions in Eqs.(\ref{eq:TotEnergyFockLowest}--\ref{eq:degen}), lead to a simple formula for the trap lift
\begin{equation}
V_1=0; \quad \Vlift=V_2=V_3 = M \Uint .
\end{equation}
It is important to note that this formula implies that it is not necessary to know the initial number of particles, $N$, to keep the well defined number $M$ of particles in the left trap, as $\Vlift$ only depends on $M$.

Let us consider the case of $N=3$ with the target state 
$\ket{\psit} = \ket{1~0~2}$, for which the energies of the Fock states with only the first two energy levels considered are shown in \fref{fig:es_3qho_3P_FockEn}(a).
One can immediately see that degeneracies appear at $\Uint = \{-\frac{1}{3}, 0, 1\}$, which correspond to drops in fidelity in the vicinity of these interaction energy values, see \fref{fig:es_3qho_3P_FockEn}(c).
As the spectrum will consist of more and more bands for increasing particle numbers, it is easy to see that the interaction region in which high-fidelity particle separation of the kind $\ket{N~0~0}\rightarrow\ket{1~0~(N-1)}$ is possible, will become more and more fragmented due to additional crossings.

If we consider the separation process which transfers a finite number of particles out of the original trap
\begin{equation}
\ket{\psii} = \ket{N~0~0}
\ \rightarrow \
\ket{\psit} = \ket{(N-M)~0~M} ,
\end{equation}
the lift required by Eqs.(\ref{eq:TotEnergyFockLowest}--\ref{eq:degen}) is now given by
\begin{equation}
V_1=0; \quad \Vlift=V_2=V_3 = (N-M)\Uint, 
\end{equation}
and depends only on the number of separated particles $N-M$. While this is a complication, for $M=1$ the lift guarantees that the SAP triplet is always energetically isolated from other Fock states in the repulsive regime.

To see this we show in \fref{fig:es_3qho_3P_FockEn}(b) the spectrum of the three-particles Fock states  for $N=3$ and $M=1$
with $\mlvl = 2$.
While there are  multiple level crossings visible in the attractive interaction regime, the energy of the SAP triplet is the lowest for any repulsive value of $\Uint$. This leads to a broad plateau in which the separation process gives high fidelities. Increasing the number of particles further leads to a denser and denser
spectrum, but the SAP band remains the lowest, and therefore isolated, over the full repulsive interacting range, allowing the separation process for a single particle in principle to work for all possible initial particle numbers.
It is worth noting though that the changes in the tunnling strengths between the traps during the dynamics of the SAP process results in non-zero bands widths and therefore potential overlaps in denser spectra. This can lead to transitions out of the SAP triplet that lower the process fidelity.
While in principle the bands can be kept arbitrarily narrow by decreasing the minimum approach distance between the traps, this would 
come at the price of having to increase the total time of the process, which is highly undesirable. 
It is therefore important to study the process in experimentally-realistic settings.

\section{Radio frequency traps}
\label{sec:rf_background}

While truncated harmonic potentials are very convenient for theoretical studies as they guarantee fulfilment of the resonance condition, they are experimentally unrealistic. We will therefore in the following examine a setup using radio-frequency (RF) traps to show that the process discussed above is viable as a quantum engineering technique.
The physics of RF traps is well studied~\cite{Zobay2001, Morgan2011, Schumm2005} and they are flexible tools that are available in many laboratories worldwide, making them ideal candidates to study the particles separation protocol.

\subsection{System}

\begin{figure}
\begin{center}
\includegraphics[width=0.8\linewidth]{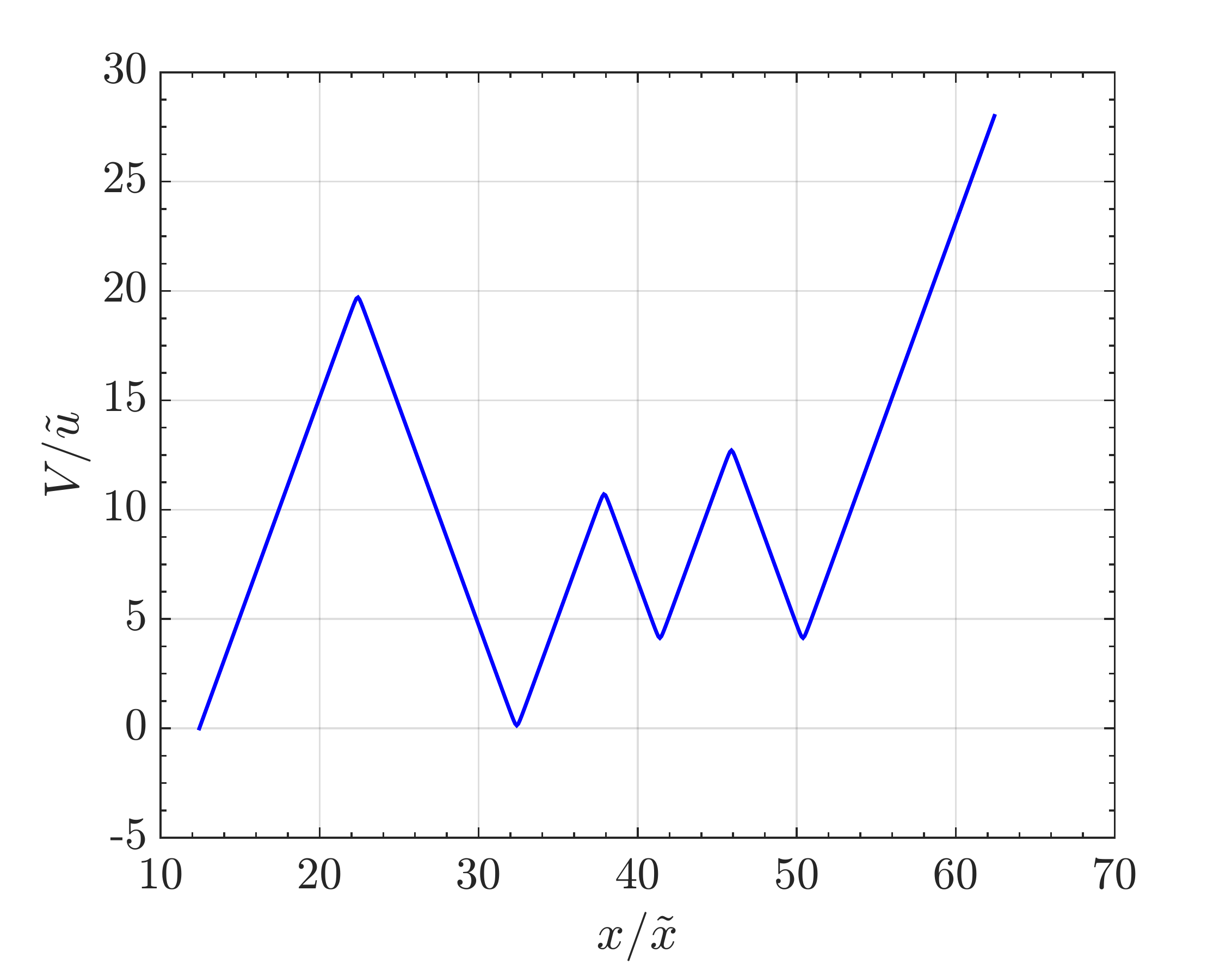}
\caption{Triple-well RF potential generated using six different frequencies and the parameters given in the text. 
The position of the left trap is fixed at $x_1 =  20\  \tilde{x}$ and the corresponding RF is $\omega_2^0 = \mu g_F b x_1 / \hbar \approx 596$ kHz. The maximum distance between the middle and the left or the right trap is $d = 9~\tilde{x}$ and
the corresponding minimum distance is $d_\textrm{min}=6~\tilde{x}$. 
The difference between frequencies at time $t=0$ is $\Delta \omega = \mu g_F b d / (2\hbar)\approx134$ kHz and the frequency for the left trap edge is $\omega_1^0 = \omega_2^0 - 2\Delta\omega\approx 328$ kHz. 
All the other frequencies are $\omega_i^0 = \omega_2^0 + (i-2)\Delta\omega,\ i=3,\dots,6$.}
\label{fig:6rf_pot}
\end{center}
\end{figure}

We consider an atom with two hyperfine sublevels ($m_F =\pm \frac{1}{2}$) in an inhomogeneous magnetic field, $B(x)=b x$, which is irradiated by a linearly polarized RF field, $\vec{B}_\mathrm{rf}\cos(\omega t)$. In this setup the atom will experience an external potential with a minimum at position $x_0$ corresponding to the resonance condition $\mu g_F m_F b x_0 = \hbar \omega$ \cite{Courteille2006}, 
where $\mu \approx9.27\times10^{-24}$ A/m$^2$ is the Bohr magneton and $g_F$ is the atomic $g$-factor. 
Using more than one RF allows to create a multi-trap potential \cite{Courteille2006} and a triple-well setup can be realised using six different frequencies \cite{Morgan2011} (see \fref{fig:6rf_pot}).

The external potential felt by the atom is then be described by~\cite{Courteille2006,Morgan2011}
\begin{equation}
V_+(x) = (-1)^{n(x)}\left[E_+(x) - \frac{\hbar\omega_{n(x)}}{2}\right]-\sum_{k=1}^{n(x)-1}(-1)^k\hbar\omega_k,
\end{equation}
where 
\begin{align}
E_+(x) &= \frac{1}{2}\sqrt{\hbar^2\Omega^2 + (\mu g_F b x - \hbar\omega_{n(x)} + 2 L_{n(x)}(x))^2},  \\
L_n(x) &= \sum_{j\neq n}\frac{\hbar^2\Omega^2}{4[\mu g_F b x - \hbar\omega_{j}]}, 
\end{align}
and $n(x)$ is chosen such that $\mu g_F b x - \hbar\omega_{n(x)}$ is minimized for all $x$, i.e., $n(x)$ is the label of the most relevant frequency at each point.
The distance between the traps and their respective ground state energies (which depend linearly on the trap lift) can be controlled by changing the $\omega_i$, which can be done with great precision. 

For our simulations we use the following experimentally realistic parameters:~the magnetic field gradient is chosen to be $b = -213$ G/cm, the atomic $g$-factor is $g_F=-\frac{1}{2}$, the Rabi frequency is $\Omega = 2\pi\times0.5$ kHz and the mass of a $^{87}$Rb atom is $\mRb=1.44\times10^{-25} \mathrm{kg}$.
For convenience we also scale all lengths by $\tilde{x}=\left({4\hbar^2}/{\mu g_F b \mRb}\right)^{\frac{1}{3}} \approx 3.18\times10^{-7}$~m, time by $\tilde{t}= \left({16\hbar \mRb}/{(\mu g_F b)^2}\right)^{\frac{1}{3}}\approx 1.34\times10^{-4}$~s, and energy by $\tilde{u}={\mRb \tilde{x}^2}/{\tilde{t}^2} \approx 7.85\times10^{-31}$~J. 
In these units, the Tonks--Girardeau regime is achieved at $\Uint^\textrm{TG}\approx 1.64\tilde{u}$.

\subsection{Particle separation}

\begin{figure}
\begin{center}
\includegraphics[width=0.8\linewidth]{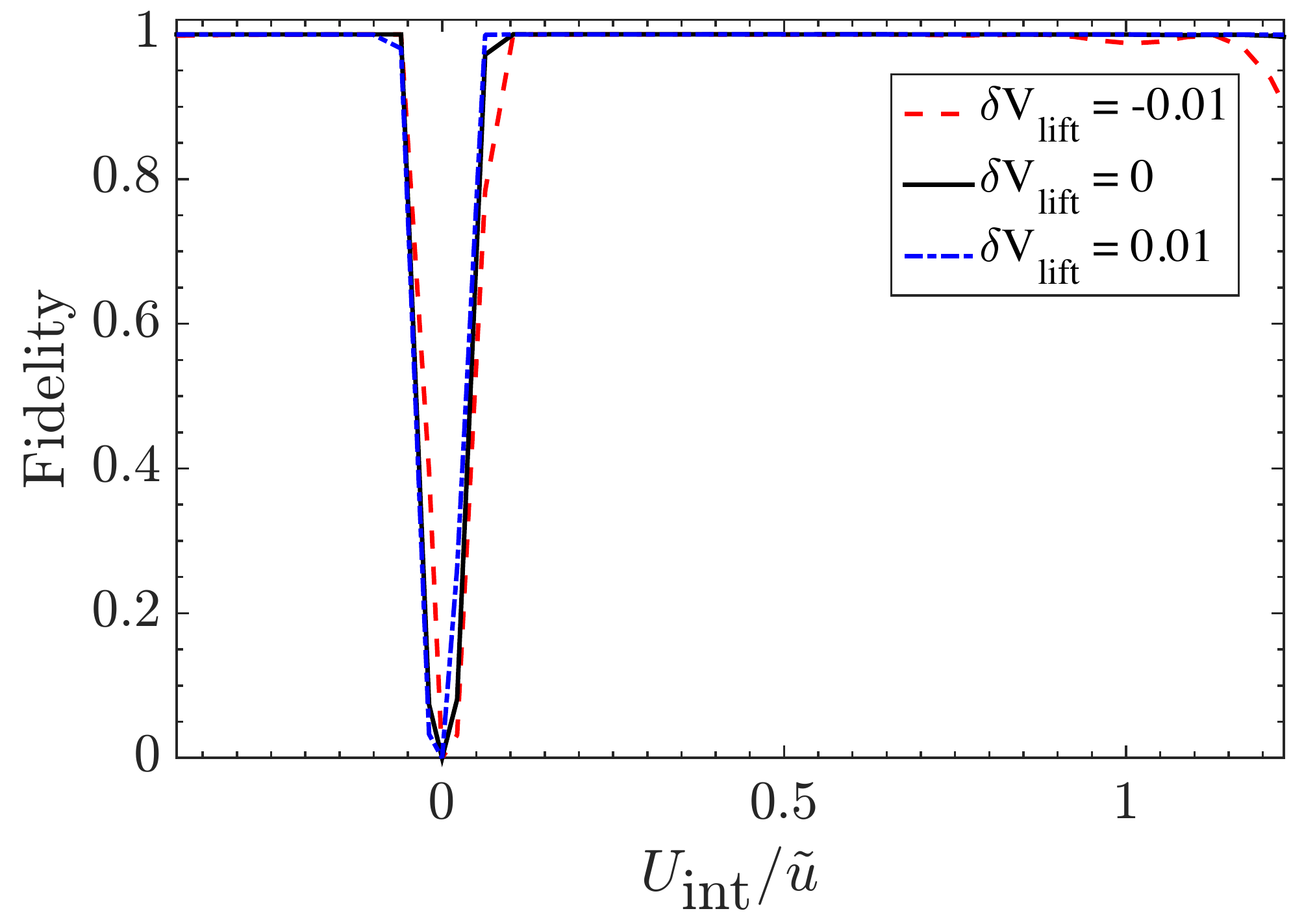}
\caption{Fidelity of the particle separation protocol in RF traps with perturbed maximum lift value.}
\label{fig:es_RF_stabil}
\end{center}
\end{figure}

In order to implement the SAP particle separation protocol, we use the following time-dependencies of the RFs
\begin{subequations}
\begin{align}
\omega_1(t) &= \omega_1^0 -\frac{\Vlift}{\hbar} \\
\omega_2(t) &= \omega_2^0 \\
\omega_3(t) &= \omega_3^0 + \frac{1}{2}f_1(t) + \frac{\Vlift}{\hbar}\\
\omega_4(t) &= \omega_4^0 + f_1(t) \\
\omega_5(t) &= \omega_5^0 + \frac{1}{2}(f_1(t) + f_2(t)) \\
\omega_6(t) &= \omega_6^0 + f_2(t),
\end{align}
\end{subequations}
where $\Delta E_1$ is the maximum lift value and
\begin{subequations}
\begin{align}
f_1(t) &= -\frac{\mu \, g_F \, b \, d_\textrm{min}}{\hbar} f(t, 0) , \\
f_2(t) &= -\frac{\mu \, g_F \, b \, d_\textrm{min}}{\hbar}  \left[f(t, 0) +f(t, \delta t)\right],\\
f(t, \delta) &=
\begin{cases}
\sin^2(\frac{2\pi (t-\delta)}{T})   & \quad 0 \leq t - \delta < \frac{T}{2} ,
\\ 
0 &\quad \text{otherwise.}
\end{cases}
\end{align}
\end{subequations}
The parameters $\delta t$ is the trap movement delay, $d_\textrm{min}$ is the minimum distance between the middle and the left or the right traps and $T$ is the total duration of the process.

For simplicity, and at variance with the truncated harmonic trap case in the previous section, where we raised the energy of the middle and right traps by $\Vlift$, here we achieve the same effect by lowering the energy of the left trap by the same amount.
This way, only one of the traps is affected by the energy shift.

We simulate the SAP separation process by starting with two atoms cooled to the ground state of the left trap with total energy $E_g$. 
Similarly to the truncated triple harmonic potential case, we calculate the energy lift value $\Vlift = \Uint = E_g - E_0$, where $E_0$ is the ground state energy of two non-interacting atoms in the left trap.
Both $E_g$ and $E_0$ are calculated numerically by direct diagonalization of the Hamiltonian.
The solid line in \fref{fig:es_RF_stabil} shows the resulting fidelities of the particle separation process and one can immediately see that for a wide range of repulsive interactions the process results in high-fidelity particle separation.

To account for possible experimental uncertainties in determining the interaction energy, we also show the particle separation fidelities for small errors in the energy shift in \fref{fig:es_RF_stabil}. 
We considered both negative and positive perturbation values $\delta\Vlift$, and the effective energy shift value used in the simulation is calculated as $V_\textrm{eff} = \Vlift + \delta\Vlift$. 
One can see that the proposed implementation of the particle separation protocol is robust against small errors in the interaction energy measurements as well as against an imperfect execution of the lowering of the left trap.
The robustness of the SAP protocol is discussed in more details in \cite{Eckert2004}.

It is important to note that the model we use to describe the RF traps is only valid when the RFs used are sufficiently far from each other~\cite{Courteille2006}. 
When the frequencies come too close, the resulting potential becomes discontinuous and does no longer describe the experimental situation. 
This limitation has been taken into account in our simulations by ensuring that the closest approach of two frequencies leads to a discontinuity smaller than $0.01\ \tilde{u}$, which has a negligible effect on the dynamics.
Furthermore,  lowering the ground state energy of the left trap requires to adjust the change of the RFs $\omega_1$ and $\omega_3$. 
While this has an effect on the middle trap, it is also very small (on the order of $0.002\ \tilde{u}$) and therefore has also no real effect on the process fidelity.

\subsection{Scaling with the number of particles}
\begin{figure}
\begin{center}
\includegraphics[width=1\linewidth,trim={2cm 6cm 3cm 7cm},clip]{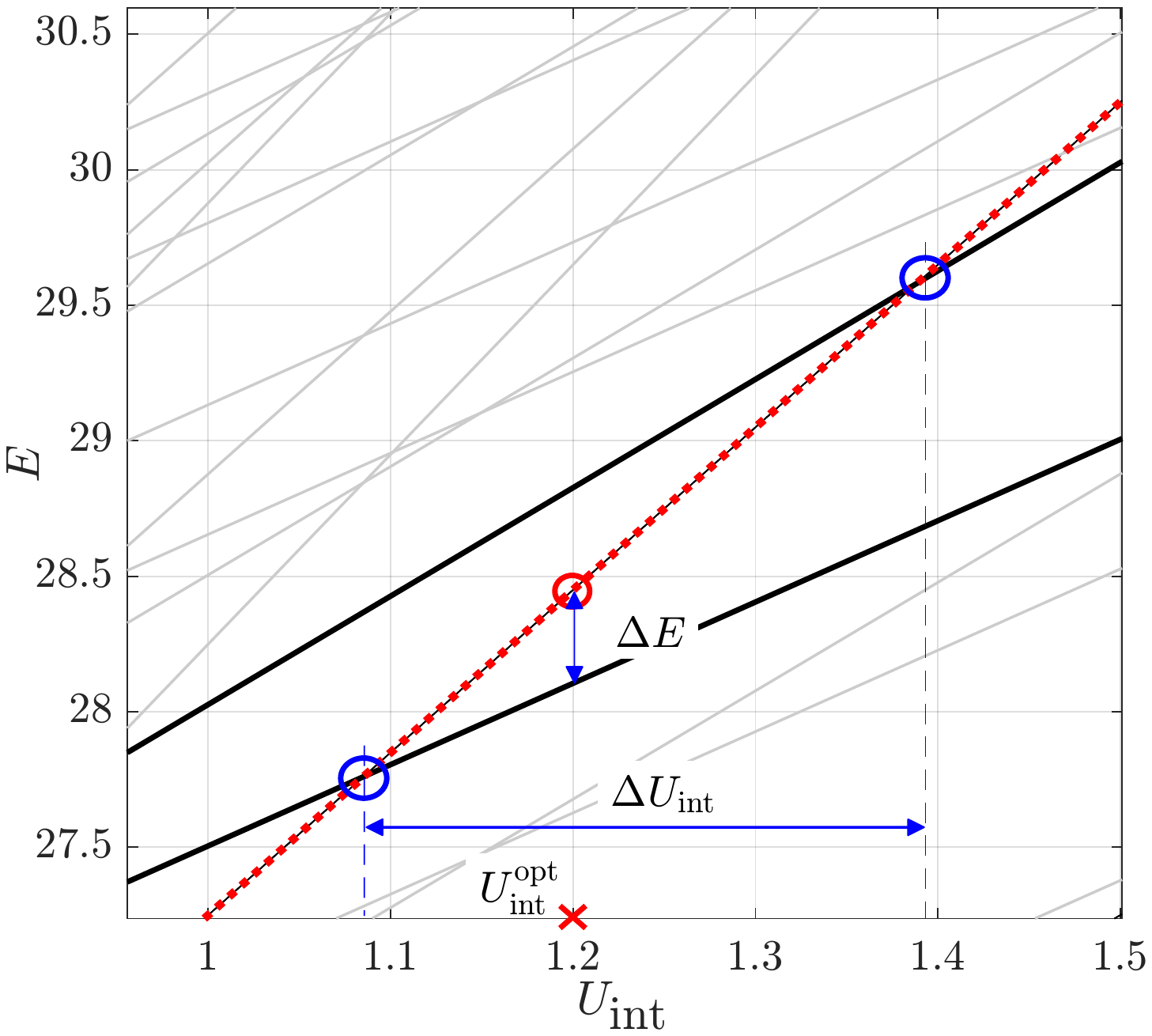}	
\caption{Schematic indicating the definitions of $\Delta E$, $\Delta \Uint$, and $\Uintopt$ using a Fock space energy spectrum.
The  dotted red line corresponds to the energy of the SAP triplet, and the black and gray lines show the energies of the other Fock states in the system (cf.~Fig.~5). 
The blue circles indicate the points of intersection of the SAP triplet energy band with the closest, neighbouring energy bands.
\label{fig:gaps}
}
\end{center}
\end{figure}

Let us finally discuss the limits of the proposed protocol. Since for increasing numbers of particles the energy spectrum becomes more fragmented, it will be harder and harder to keep the system within the SAP triplet. 
To quantify the limit we determine the size of the maximum energy gap, $\Delta E$, between the SAP triplet and the neighboring bands as a function of  the initial and final number of particles in the left trap 
over the whole range of repulsive interactions. In addition we define the value of the interaction strength corresponding to this maximal energy gap, $\Uintopt$, and the distance between the two points at which the SAP triplet crosses other bands, $\Delta \Uint$  (see \fref{fig:gaps}).
Since at the point $\Uintopt$ the process works best, it gives a good insight for its limits: 
1) $\Delta E$ quantifies how hard it is to follow the SAP state
and 2) $\Delta \Uint$ indicates how fragmented the region, in which a high fidelity process can be expected, has become.

\begin{figure}
\begin{center}
\includegraphics[width=1\linewidth,trim={1cm 1cm 1cm 1cm},clip]{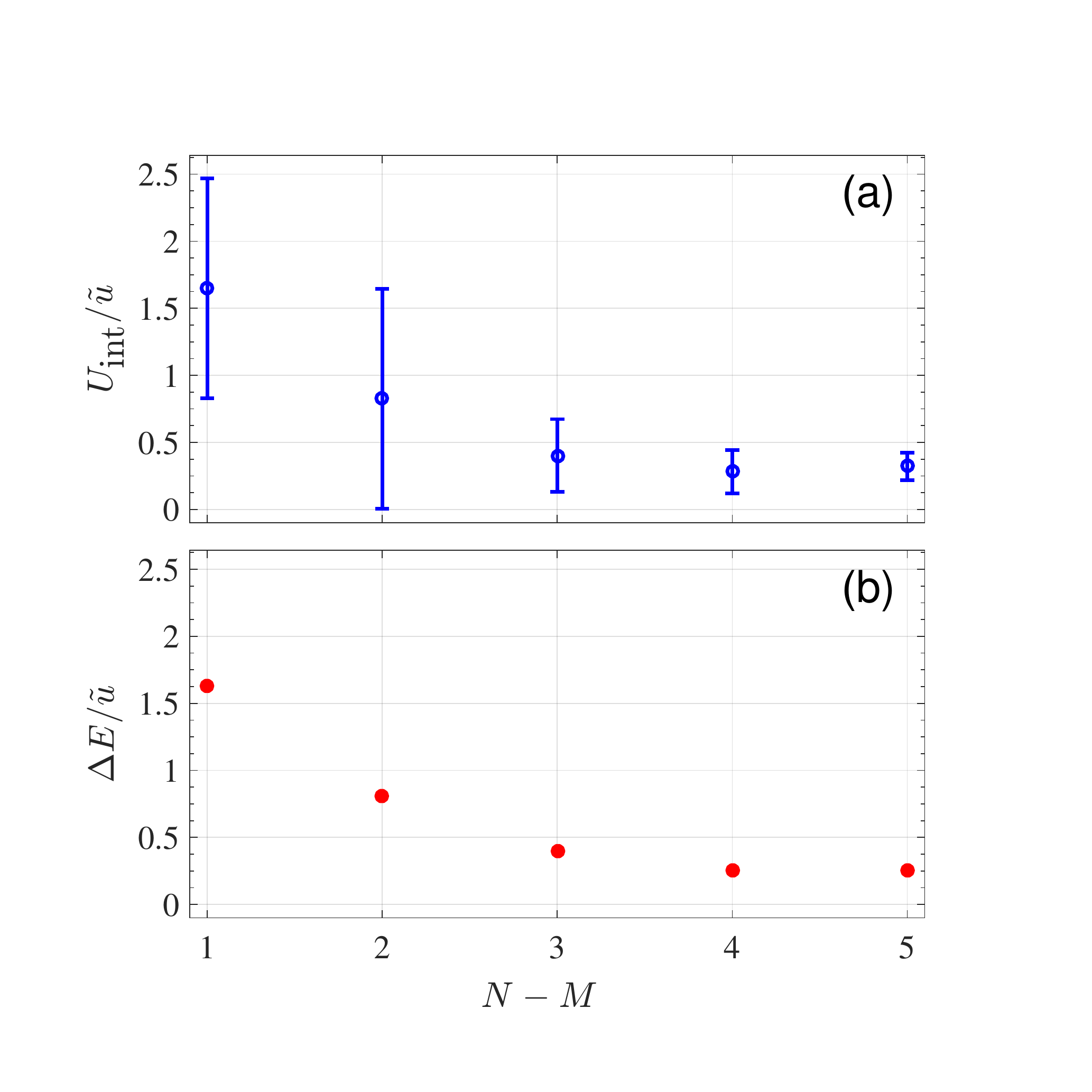}	
\caption{(a) Optimal interaction energy value $\Uintopt$ with its margin $\Delta \Uint$ as an error bar and (b) energy gap $\Delta E$.
The Fock energy calculation are performed for RF traps using the Bose--Hubbard model for the three lowest energy bands.
\label{fig:scaling}
}
\end{center}
\end{figure}

The results reveal that the value of $\Delta E$ does not depend on the initial number of particles $N$, but only on the numbers of particles that are to be moved out of the trap, $N-M$. This can be easily understood by considering the structure of the energy spectrum (see \fref{fig:es_3qho_3P_FockEn}).
The energy band of each Fock state increases linearly with the interaction energy, intersecting $\Uint=0$ at points that  correspond to combinations of the excited trap eigenstates.
The slope then depends on the number of particles in the right and the middle traps and
it is easy to see that the $N-M=1$ energy band has the smallest slope, the $N-M=2$ has the second smallest slope, etc.
With increasing $N-M$ one needs, of course, to include more energy levels in the model to account for all intersections, but the structure does not depend on $N$.
However, $\Delta \Uint$ rapidly decrease with increasing $N-M$ and in \fref{fig:scaling} we show our figures of merit. From there one can estimate that a realistic upper limit on $N-M$ is on the order of 10 particles, independent on the initial number of particles.
Thus, for any $N$, by using, for example, Feshbach resonance~\cite{Tiesinga1993, Inouye1998, Bauer2009}, one can tune the interaction energy to access the region where the particle separation protocol is the easiest.

\section{Conclusions}
\label{sec:conclusions}

In this work we have proposed a protocol based on the spatial adiabatic passage technique that allows to divide a sample of interacting particles in a controlled way. 
The technique is based on engineering a quasi-three level system  by raising or lowering the energies of some of the traps and allowing for an adiabatic transition between initial and target states.
We have explicitly examined the cases  $\ket{2~0~0}\rightarrow\ket{1~0~1}$ for a two-particle system and $\ket{3~0~0}\rightarrow\ket{2~0~1}$ for a three-particle system and shown that the SAP protocol results in high-fidelities over large ranges of interaction energies. 
The regions where the protocol fails can be found from the level crossings present in the spectrum of a Bose--Hubbard model.

We have also shown that this protocol is realistic and robust against experimental uncertainties by examining a setting where two $^{87}$Rb atoms were trapped in a radio-frequency trap setup.
Using experimentally realistic parameters, the same high fidelities were obtained as for the idealised system, showing that quantum engineering techniques based on spatial adiabatic passage are useful for interacting particle systems.
The protocol we proposed is independent of the number of initial particles and  can be therefore be used also in systems with large initial particle numbers.

\section{Acknowledgments}
This project was supported by the Okinawa Institute of Science and Technology Graduate University.

\appendix

\section{Tunneling couplings}

The tunneling term in the BH Hamiltonian~\eqref{eq:BHHam} is defined as
\begin{align}
H_\text{tunnel} &=
\sum_{i=1}^2
\sum_{N^\textrm{trap}_i=1}^{N}
\sum_{N^\textrm{trap}_{i+1}=0}^{N-N^\textrm{trap}_i}
\sum_{p=1}^{N^\textrm{trap}_i}
\sum_{\substack{  \vec{M}\in\mathcal{P}(N^\textrm{trap}_i-p) \\ \vec{p}\in\mathcal{P}(p)}}  \sum_{\substack{\vec{K}\in\mathcal{P}(N^\textrm{trap}_{i+1}) \\ \vec{q}\in\mathcal{P}(p)}} \nonumber \\
 &\qquad \left(\Omega_{i}^{\vec{p}\vec{q}}(\vec{M}, \vec{K})\prod\limits_{j=0}^{\mlvl-1}\hat{a}_{i+1~j}^{\dagger q_j}\hat{a}_{i~j}^{p_j}  + h.c.\right)
 \label{eq:Htunnel}
\end{align}
and it includes all tunnelling events of $p$ particles between the traps $i$ and $i+1$. The set $\mathcal{P}(n)$ contains all possible ways to distribute $n$ particles into $\mlvl$ energy levels of one trap, and
\begin{equation}
\sum_{j=0}^{\mlvl-1}p_j = \sum_{j=0}^{\mlvl-1}q_j = p.
\end{equation}
The corresponding coupling coefficients $\Omega_{i}^{\vec{p}\vec{q}}(\vec{M}, \vec{K})$ denote the tunnelling frequencies of $p$ atoms between the level occupation configurations $\vec{p}=\left(p_0,\ldots,p_{\mlvl-1}\right)$ and $\vec{q}=\left(q_0,\ldots,q_{\mlvl-1}\right)$ of traps $i$ and $i+1$ respectively.

In what follows we derive the tunneling coupling amplitudes between two general Fock states 
\begin{equation}
\ket{\psii}_i^{\vec{p}\vec{q}}=\bigket{\begin{smallmatrix}
						M_{0i} + p_0&K_{0(i+1)}&\\
						M_{1i}+p_1&K_{1(i+1)}&\\
						\ldots&\ldots&\\
						M_{(\mlvl-1)i}+p_{\mlvl-1}&K_{(\mlvl-1)(i+1)}&
					\end{smallmatrix}}
\end{equation} and
 \begin{equation}
 \ket{\psit}_i^{\vec{p}\vec{q}}=\bigket{\begin{smallmatrix}
						M_{0i} &K_{0(i+1)}+ q_0&\\
						M_{1i}&K_{1(i+1)}+q_1&\\
						\ldots&\ldots&\\
						M_{(\mlvl-1)i}&K_{(\mlvl-1)(i+1)}+q_{\mlvl-1}&
					\end{smallmatrix}},
\end{equation}
which contain occupation numbers for traps $i$ and ${i+1}$. 
The coupling coefficient between these two states is defined from the general Hamiltonian~\eqref{eq:ham2p} as
\begin{equation}
\Omega_{i}^{\vec{p}\vec{q}}(\vec{M}, \vec{K}) =\opbraket{\psit}{\hat{H}}{\psii}_i^{\vec{p}\vec{q}}.
\label{eq:couplings}
\end{equation}
If $\Omega_{i}^{\vec{p}\vec{q}}(\vec{M}, \vec{K})\neq 0$, then the corresponding relevant term that will appear in the BH Hamiltonian is proportional to $\prod\limits_{j=0}^{\mlvl-1}\hat{a}_{i+1~j}^{\dagger q_j}\hat{a}_{i~j}^{p_j}$, thus
\begin{equation}
\begin{split}
\Omega_{i}^{\vec{p}\vec{q}}(\vec{M}, \vec{K}) &= \tilde{\Omega}_{i}^{\vec{p}\vec{q}}\opbraket{\psit}{\prod\limits_{j=0}^{\mlvl-1}\hat{a}_{i+1~j}^{\dagger q_j}\hat{a}_{i~j}^{p_j}}{\psii}_i^{\vec{p}\vec{q}}\\
&=\prod\limits_{j=0}^{\mlvl-1}\sqrt{\frac{(M_j+p_j)!}{M_j!}\frac{(K_j + q_j)!}{K_j!}}\tilde{\Omega}_{i}^{\vec{p}\vec{q}}.
\end{split}
\end{equation}
If $\vec{M}=\vec{0}$ and $\vec{K}=\vec{0}$, then
\begin{equation}
\Omega_{i}^{\vec{p}\vec{q}}(\vec{0}, \vec{0})  = \prod\limits_{j=0}^{\mlvl-1}\sqrt{p_j!q_j!} \tilde{\Omega}_{i}^{\vec{p}\vec{q}},
\end{equation}
where $\tilde{\Omega}_{i}^{\vec{p}\vec{q}}$ is the tunnelling frequency of all $p$ atoms between level occupation configurations $\vec{p}$ of the trap $i$ and $\vec{q}$ of the empty trap $i+1$.

Since only the order of magnitude is important in order to show the shape of the regions of high-fidelity particle separation, we assume $\Omega_{i}^{\vec{p}\vec{q}}(\vec{0}, \vec{0})\approx\tilde{\Omega}_{i}^{p}$.
Here $\tilde{\Omega}_{i}^{p}$ is the tunnelling frequency of $p$ atoms between the ground states of traps $i$ and $i+1$ in the absence of other atoms.
In the three particle calculations we assumed $\tilde{\Omega}_{i}^{3}\propto\tilde{\Omega}_{i}^{2}$, while $\tilde{\Omega}_{i}^{2}$ and $\tilde{\Omega}_{i}^{1}$ were calculated numerically.
\Eref{eq:couplings} can thus be written as
\begin{equation}
\Omega_{i}^{\vec{p}\vec{q}}(\vec{M}, \vec{K}) \approx \prod\limits_{j=0}^{\mlvl-1}\sqrt{\frac{(M_j+p_j)!}{M_j!}\frac{(K_j + q_j)!}{K_j!}\frac{1}{p_j!q_j!}}\tilde{\Omega}_{i}^{p}.
\end{equation}

\end{document}